\documentclass[useAMS,usenatbib]{mn2e}
\usepackage{epsfig}

\def\ifm#1{\relax\ifmmode#1\else$\mathsurround=0pt #1$\fi}
\def\msun{{\rm M}_{\odot}}

\def\ltsima{$\; \buildrel < \over \sim \;$}
\def\lsim{\lower.5ex\hbox{\ltsima}}
\def\gtsima{$\; \buildrel > \over \sim \;$}
\def\gsim{\lower.5ex\hbox{\gtsima}}

\def\oiilam{[O\,{\scriptsize II}]\,$\lambda3727$}
\def\oiii{[O\,{\scriptsize III}]}
\def\oiiilam{[O\,{\scriptsize III}]\,$\lambda\lambda4959,5007$}
\def\feii{Fe\,{\scriptsize II}}
\def\feiilam{Fe\,{\scriptsize II}\,$\lambda\lambda4434$-$4684$}

\def\mgiilam{Mg\,{\scriptsize II}\,$\lambda2800$}
\def\hb{H$\beta$}
\def\simlt{\lower.5ex\hbox{$\; \buildrel < \over \sim \;$}}
\def\simgt{\lower.5ex\hbox{$\; \buildrel > \over \sim \;$}}
\def\kms{\mbox{ km s$^{-1}$}}
\def\mpc{\mbox{ Mpc}}
\def\kpc{\mbox{ kpc}}
\def\pc{\mbox{ pc}}
\def\Ho{\mbox{H}_0}
\def\mag{\mbox{ mag}}
\def\arcdeg{$^{\circ}$}

\newcommand\araa{ARA\&A}

\begin{document}

\title[Spectroscopic signature of lensing dark matter
substructure]{Detecting dark matter substructure spectroscopically in
strong gravitational lenses}

\author[Moustakas \& Metcalf]
{Leonidas~A.~Moustakas\,$^1$ \& R. Benton Metcalf\,$^2$\\
$^1$Astrophysics, University of Oxford, Keble Road, Oxford, OX1\,3RH,
UK\thanks{\tt leonidas@astro.ox.ac.uk}\\ 
$^2$Institute of Astronomy, University of Cambridge, Madingley Road,
Cambridge, CB3\,0HA, UK\thanks{\tt bmetcalf@ast.cam.ac.uk}} 

\date{Submitted 2002 May 31}

\maketitle

\begin{abstract}
The Cold Dark Matter (CDM) model for galaxy formation predicts that a
significant fraction of mass in the dark matter haloes that surround
$L\sim L_*$ galaxies is bound in substructures of mass
$10^4$--$10^7\,\msun$.  The number of observable baryonic
substructures (such as dwarf galaxies and globular clusters) falls
short of these predictions by at least an order of magnitude.  We
present a method for searching for substructure in the haloes of
gravitational lenses that produce multiple images of QSOs, such as
4-image Einstein Cross lenses.  Current methods based on broadband
flux ratios cannot cleanly distinguish between substructure,
differential extinction, microlensing and, most importantly,
ambiguities in the host lens model.  These difficulties may be
overcome by utilizing the prediction that when substructure is
present, the magnification will be a function of source size.  QSO
broad line and narrow line emission regions are approximately $\sim
1$\,pc and $>100$\,pc in size, respectively.  The radio emission
region is typically intermediate to these and the continuum emission
region is much smaller.  When narrow line region (NLR) features are
used as a normalisation, the relative intensity and equivalent width
of broad line region (BLR) features will respectively reflect
substructure-lensing and microlensing effects.  Spectroscopic
observations of just a few image pairs would probably be able to
cleanly extract the desired substructure signature and distinguish it
from microlensing -- depending on the {\em actual} level of projected
mass in substructure.  In the rest-optical, the \hb/\oiii\, region is
ideal, since the narrow wavelength range also largely eliminates
differential reddening problems.  In the rest-ultraviolet, the region
longward and including Ly$\alpha$ may also work.  Simulations of
Q2237+030 are done as an example to determine the level of
substructure that is detectable in this way, and possible systematic
difficulties are discussed.  This is an ideal experiment to be carried
out with near-infrared integral field unit spectrographs on 8-m class
telescopes, and will provide a fundamentally new probe of the internal
structure of dark matter haloes.
\end{abstract}


\section{Introduction\label{sec:intro}}

Galaxy formation models cast in a $\Lambda$CDM universe have
experienced considerable success to date.  With a fixed choice of
cosmological parameters, it is possible to make specific predictions,
for example based on N-body simulations, regarding the structure 
of individual dark matter haloes.  One of the challenges in
this field is to calculate the level of the continuously-infalling
substructure that will survive within large haloes.  Present N-body
simulations indicate that there should be plentiful substructure in a
$L_*$ galaxy's halo, with masses above $10^7\msun$; below this scale
limitations in resolution becomes important.
If these subhaloes contain stars, there should be significantly more
dwarf galaxies near the Galaxy than are seen \citep{moore:99,
klypin:01}.  Various solutions to this excess problem have been
proposed, including warm dark matter (which erases substructure on
small scales; e.g. \citealt{avila:01}); self-interacting dark matter
(which causes subhaloes to evaporate; Spergel \& Steinhardt 2000); and
radiation-driven ionization of the intergalactic medium such that star
formation in dwarf-size galaxies is suppressed at early epochs, as
discussed in \citet{somerville:02}, \citet*{bullock:01}, and
\citet{klypin:99}.  If the substructure does exist, its gravitational
influence should be detectable, and therefore this prediction of
$\Lambda$CDM in particular may be tested.  There is already some
evidence in support of the need for substructure to explain the
relative magnifications in strong lenses; disentangling its signature
from other (equally interesting) effects is the scope of this paper.

From the positions of the images and lens galaxy of a 4-image lens
system, a parametric smooth mass model for the lensing galaxy and halo
can be computed (including external shear), along with predictions for
the three independent flux ratios.  Measuring deviations between the
predictions and the (differential-reddening-corrected) observed
integrated fluxes in real Einstein Cross lenses, provides a direct
estimate of the surface density in substructure
\citep*{metcalf:01,chiba:02,metcalf:02,dalal:02}.  Substructure has
long been considered a possible explanation for inconsistencies in the
observed flux ratio of B1422$+$231 \citep*{mao:98}.  However, this
approach has a vital flaw.  The predicted flux ratios can be strongly
dependent on the parametric lens models used, making any measurement
of the mass fraction in substructure \cite[e.g.][]{dalal:02} suspect.
The observational technique proposed here largely avoids this
important problem, as well as the problems of intrinsic variability
(on the timescale of the typical time-delay between images), and the
effects of microlensing, with the possible change of continuum slope
as well as intensity.

In \S\ref{sec:models} the main features of substructure and its
lensing effects are summarized.  Simulations demonstrating the
importance of source-size on magnification are presented in
\S\ref{sec:lensing-simulations}, for a specific lens, Q2237+030.  The
spectroscopic approach proposed here is discussed in \S\ref{sec:spec},
with the most difficult problems and their possible solutions
addressed.  In \S\ref{sec:disc} we discuss selection criteria for lens
candidates for this experiment, as well as modes of observations that
would be sufficient for obtaining results.  The main conclusions are
summarized in \S\ref{sec:conc}.  In the Appendix an estimate of the
observational uncertainties is calculated.  All cosmological
calculations are carried out for a $\Lambda$CDM cosmology, with
$\Omega_{\Lambda}=0.7$, $\Omega_0=0.3$, though none of the arguments
(except the level of substructure one might expect) depend strongly on
this choice.  The Hubble parameter is $\Ho= 65\,h_{65}\kms\mpc^{-1}$.

\section{Substructure \& Gravitational Lens Models}\label{sec:models}

\subsection{substructure and small-mass dark matter haloes}

The highest-resolution N-body simulations of structure formation are
currently limited to $\sim 10^6\msun$ particles
\citep[e.g.][]{moore:99}.  Down to the scale of 10 particles
($\sim10^7\msun$) there is clear evidence for surviving substructure
within large dark matter haloes with a power-law mass function, $dN/dm
\propto m^{-\alpha}$, where $\alpha\simeq 1.91$.  For observed dwarf
galaxies, $\alpha \simeq 2.35$ \citep{klypin:01}.  In these
simulations there is a continuous process of infall and merging
working against tidal stripping and dynamical friction to maintain the
substructure population.  Within the virial radius of a galactic halo,
about $10$--$15$\% of the mass is in structure of mass $\simgt
10^7\msun$.  Analytic arguments suggest that future simulations with a
larger dynamic range will find that substructure survives down to far
smaller masses than are currently accessible by direct computation
\citep*{metcalf:01}.

In addition to the substructure inside the primary lens, there will
also be small haloes in intergalactic space in front of and behind the
primary lens.  Zhao \& Metcalf (2002; in prep.) show that the
importance of these two populations to compound lensing is
approximately equal, if the tidal destruction of haloes within the lens
is negligible.  Tidal destruction will increase the relative
importance of intergalactic haloes although the importance of this
cannot yet be accurately quantified.  Despite this, in our estimates
and simulations we treat all the haloes as if they are at the same
redshift as the primary lens.  Since the strength of a lens of fixed
mass reaches a maximum when the lens is half way to the source,
measured in angular size distance, and the primary lens is usually
near this point, an estimate of the surface density in substructure
under this simplification will be an estimate of the minimum density
required to produce the observed lensing effects.

The importance of haloes to lensing is strongly dependent on their
internal structure.  To quantify the internal structure in some of the
estimates given below, we use the parameter $\gamma(r) \equiv d\ln
M(r)/d\ln r$, where $M(r)$ is the mass of the halo within a projected
radius of $r$.  The tidal radius of a subhalo is set by the
requirement that the average density within that radius be
proportional to the average density of the host halo within the orbit
of the subclump -- the constant of proportionality depends on the
structure of the host.  This means that the average density
$\overline{\rho}_c = 3m/4\pi r(m)^3$, where $r(m)$ is the radius of a
subhalo, will be independent of mass.  Additionally, in the
hierarchical structure formation model, the initial density of a halo
is proportional to the average density of the universe at the time
when the halo first collapsed.  In the CDM model all the small-mass
haloes of interest here are almost coeval.  This results in the same
conclusion for the intergalactic haloes as for the subhaloes --
$\overline{\rho}_c$ is independent of mass although it may have a
different value for the two populations.  We will take
$\overline{\rho}_c$ to be a free parameter in our discussions.

\subsection{the influence of source size on lensing}
\label{sec:infl-source-size}

We can make some estimates for how compound lensing will be affected
by the size of the source QSO using the simple model of substructure
outlined above.  Let us brake the magnification matrix, $A_{ij}$, into
a contribution from the smooth lens and a perturbation caused by the
small clumps -- $A_{ij}=A^o_{ij}+\delta A_{ij}$.  Let us say a clump
will have a significant influence on the magnification if $\delta
A_{ij}$ is of order $\epsilon$ or larger.  The value of $\epsilon$
will depend on the magnification caused by the unperturbed smooth lens
($\delta\mu/\mu\sim\mu_o\epsilon$ where $\mu_o$ is the smooth lens
magnification).  The magnification of a source with finite size is the
point-source magnification integrated over the area of the image.
Because of this, a finite source sees a magnification that is smoothed
on the scale of its own size.  For a halo to have an influence, the
region where its contribution to the magnification matrix is greater
than $\epsilon$ must be larger than the image of the source.  If we
take the physical size scale of the source to be $\ell_s$, this
requirement puts a lower limit on the mass of haloes that can be
important for lensing:
\begin{eqnarray}\label{eq:m_c}
m_c= \left[ \left( \frac{\epsilon\pi\Sigma_c}{1+\gamma}\right) \left(
\frac{D_l\ell_s}{D_s}\right)^{2-\gamma} \left( \frac{4\pi}{3} 
\overline{\rho}_c\right)^{-\gamma/3} \right]^{3/(3-\gamma)}. 
\end{eqnarray}
where the critical density is $\Sigma_c=(4\pi G D_l D_{ls}/c^2
D_s)^{-1}$ and the angular size distances to the lens, source, and
from the lens to the source are denoted by $D_l$, $D_s$, and $D_{\rm
ls}$, respectively.  The logarithmic slope $\gamma$ is evaluated at
the radius where $\delta A(r) \simeq \epsilon$.

It is useful to put some typical numbers into Eqn.~\ref{eq:m_c}.
Consider a source at $z_s=2$ and a lens at $z_l=0.5$.  In this case
$\Sigma_c= 1.93\times 10^9 h_{65} \msun/{\rm kpc}^2$ and
$D_l/D_s=0.73$.  For a point-mass (the most condensed possible halo),
$\gamma=0$ and
\begin{equation}\label{mc_ps}
m_c \simeq 3.2\times 10^3 \epsilon h_{65}
\left(\frac{\ell_s}{\pc}\right)^2 \msun~~~\mbox{(point-mass lens)}. 
\end{equation}

\begin{figure}
\vspace{1.7in}
\includegraphics{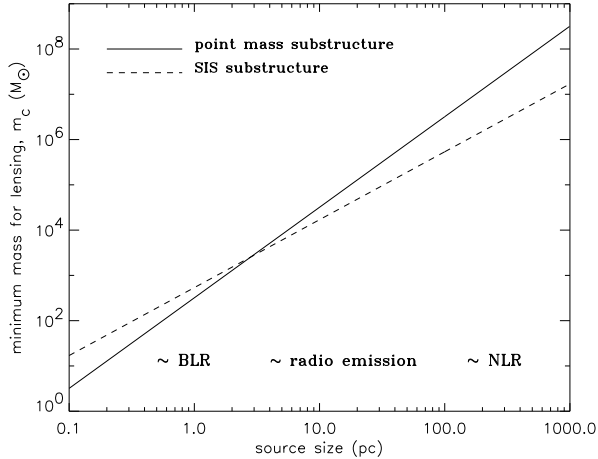}
\vspace{1in}
\caption[Minimum masses for lensing]{\footnotesize The minimum mass of
a small halo that can have an influence on the compound lensing as a
function of the source size.  The solid line is from Eqn.~\ref{mc_ps}.
The dashed line is Eqn.~\ref{mc_sis} with $R=1\kpc$ and
$\sigma_h=200\kms$.  In both cases $h_{65}=1$ and $\epsilon=0.1$.  If
the magnification is high ($\mu\gsim 10$), $\epsilon=0.01$ would be
more appropriate.  The general size range for different emission
regions of a QSO are labelled (BLR -- broad line region; NLR -- narrow
line region).  The central accretion disc of a QSO which is believed
to be the source of the continuum emission has a scale of
$10^{-4}-10^{-3}\pc$; off the scale to the left.}
\label{fig:mcls}
\end{figure}

For less condensed haloes we must make an assumption about their
internal density, $\overline{\rho}_c$.  The characteristic density of
the intergalactic haloes, $\overline{\rho}_c$, is set by the process
of structure formation at high redshift.  If these haloes fall into a
host lens they will remain intact until the average density of the
halo within their orbit, $\overline{\rho}_h(R)$, becomes of order the
density of the subhalo, then it will be tidally stripped.  If all the
subhaloes are tidally truncated, their density can be calculated from
the average interior density profile of the host,
$\overline{\rho}_h(R)$.  If the host is modeled as a singular
isothermal sphere (SIS), $\overline{\rho}_c=2 \overline{\rho}_h(R)$.
This results in a cutoff mass of
\begin{eqnarray}\label{mc_sis}
m_c \simeq 1.7\times 10^4 \msun 
\left(\epsilon h_{65}\frac{\ell_s}{\rm pc} \right)^{3/2}
\frac{200}{\sigma_h} \frac{\rm km}{{\rm s}} \frac{R}{\rm kpc} ~
\mbox{(SIS)}, 
\end{eqnarray}
where $\sigma_h$ is the velocity dispersion of the host lens.  The
haloes could be denser than assumed here if they were born denser, or
they could be less dense if they were born less dense and primarily
reside outside of the primary lens.  Since no accurate simulations of
structure formation at these small mass scales have been done, we will
take the above estimates as benchmarks for our discussions.

\begin{figure}
\begin{center}
\vspace{1.7in}
\includegraphics{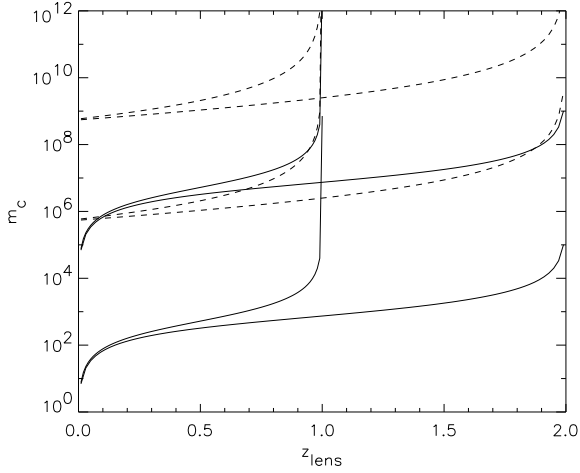}
\vspace{1in}
\caption[]{\footnotesize The minimum lensing mass as a function of the
source size as defined in Eqn~\ref{eq:m_c}.  The solid curves are for
point mass substructures and the dashed are for singular isothermal
spheres.  The lower of each set of curves are for a source size of
$\ell_s=1\pc$ and the upper for $\ell_s=1\pc$.  The curves that stop
at $z_l=1$ are for source redshift $z_s=1$.  The others are for
$z_s=2$.  The SIS are tidally truncated with $R=1\kpc$. A realistic
case would probably lie somewhere between the point mass and the SIS
curves of each type.  In all cases, $h_{65}=1$ and $\epsilon=0.1$.
\label{fig:mcofz}}
\end{center}
\end{figure}

The estimates of $m_c$ are plotted in Figure~\ref{fig:mcls} as a
function of source size along with the approximate size scales for
different emission regions of a QSO as discussed in
section~\ref{subsec:qsos}.  The image of a source of a given size will
be affected by halo masses above this limit.  For example, the image
of a 1\,pc source will reflect the influence of haloes with $m\gsim
10^3\msun$ while the image of a 100\,pc source will only be affected
by haloes with $m\gsim 10^6\msun$.  For source sizes below $\sim
10^{-2}\pc$ (such as the accretion disc of a QSO), $m_c$ is less than
a solar mass, making microlensing by ordinary stars possible.
Figure~\ref{fig:mcofz} shows $m_c$ as a function of lens redshift.
The drop in $m_c$ near $z_l=0$ in Figure~\ref{fig:mcofz} for the point
masses is somewhat misleading here because the Einstein radius becomes
very small near $z_l=0$ making these cases extremely rare.  The value
of $m_c$ for any realistic substructures lies somewhere between its
value for point mass lenses and SIS lenses.

This mass cutoff alone indicates that the magnification should be a
strong function of source size if there is a large population of
haloes with $m\lsim 10^7\msun$.  There are other factors that
influence how magnification will depend on source size.  Larger
sources are more likely to be eclipsed by rarer, larger haloes.  On
the other hand, if the haloes have high density cusps in their centers
-- as expected from simulations -- smaller sources can reach higher
magnifications.  To determine accurately how the magnification
distribution will depend on source size and halo mass distribution
requires numerical simulations.

\subsection{lensing simulations: Q2237+030}
\label{sec:lensing-simulations}

We take the image positions and lens galaxy position of the 4-image
QSO lens system Q2237+030 and fit a elliptical power-law plus shear
model to them for the smooth lens component (see \citealt{metcalf:02}
for details).  To calculate the magnifications with substructure, an
improved version of the ray tracing code used by \cite{metcalf:01} is
used.  The lensing equation is solved to obtain the source position
for each point on a coarse grid surrounding the unperturbed image
position.  The closest grid point to image position is found and a new
smaller grid is set up around that point.  This process is repeated
until the image takes up a large fraction of the grid and the desired
accuracy is reached (usually $\delta\mu/\mu =10^{-3}$).  This
grid-refinement technique allows us to achieve the large dynamical
range required to calculate the magnifications of sources with sizes
ranging from less than 1\,pc to several hundred pc while keeping the
substructure fixed.

The substructure is modeled as singular isothermal spheres truncated
such that $\overline{\rho}_c=6.9\msun\pc^{-3}$ and with a mass
function of $dN/dm \propto m^{-2}$.  The total surface density in
substructures with $10^3\msun < m < 10^6\msun$ is fixed at 10\% of the
total surface density according to the smooth model.  Random
realizations of the masses and positions of the subclumps are made and
the magnifications with several different sources are calculated in
each one.

\begin{figure}
\begin{center}
\vspace{1.7in}
\includegraphics{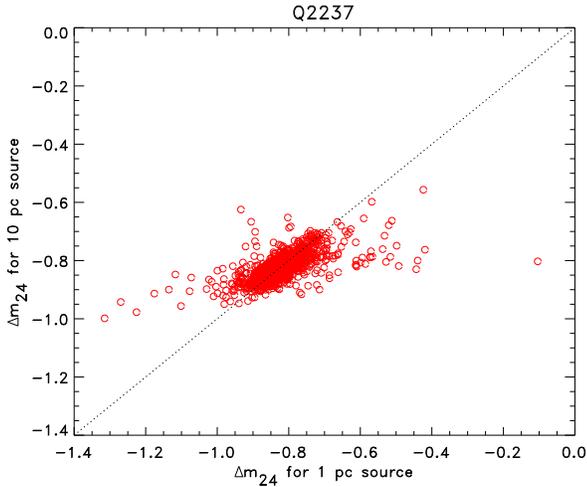}
\vspace{1in}
\caption[]{\footnotesize Simulated magnification ratios between image
B and image D of quad-lens Q2237+030 using two different source sizes.
Each circle is a different realization of the substructure.  There are
1\,000 realizations shown.  Without substructure, $\Delta m_{24}$
should not depend on source-size in this range so all the points would
be on the 45\arcdeg\ line shown.  The range of ratios along the
45\arcdeg\ line is larger, but excursions in this direction can be
degenerate with the smooth lens model.  Excursions perpendicular to
the line unambiguously indicate small-scale structure.
\label{fig:relmag}}
\end{center}
\end{figure}

\begin{figure}
\begin{center}
\vspace{1.7in}
\includegraphics{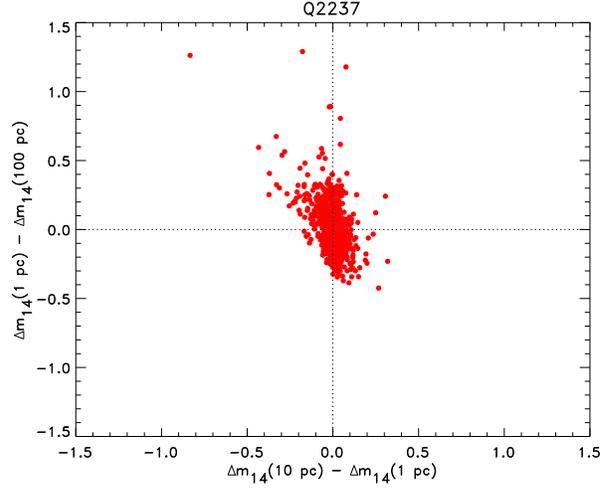}
\vspace{1in}
\caption[]{\footnotesize The differences in magnification ratios
between image A and image D of quad lens Q2237+030 using three
different source sizes (1\,pc, 10\,pc and 100\,pc).  A smooth,
substructure-free lens would predict that the three magnification
ratios would be very close to equal, the center of the plot.  These
are the same 1\,000 realizations as in Figure~\ref{fig:relmag}.  One
quarter of the realizations have $|\Delta m_{14}(10\pc)- \Delta
m_{14}(1\pc)| > 0.05\mag$ and one quarter of them have $|\Delta
m_{14}(1\pc)- \Delta m_{14}(100\pc)| > 0.18\mag$.
\label{fig:relmag_1}}
\end{center}
\end{figure}

Figure~\ref{fig:relmag} shows the flux ratio, measured in magnitudes,
of image A to image D (marked `1' and `2') in Q2237+030 for 1\,000
realizations of the substructure.  The ratios for source sizes of
1\,pc and 10\,pc are shown.  If there were not a substantial number of
haloes with mass less than $\sim 10^6\msun$, we would expect these two
ratios to be equal and the circles in the plot to lie on the diagonal
line.  No smooth model can account for points being off this line.

\begin{figure}
\begin{center}
\vspace{1.7in}
\includegraphics{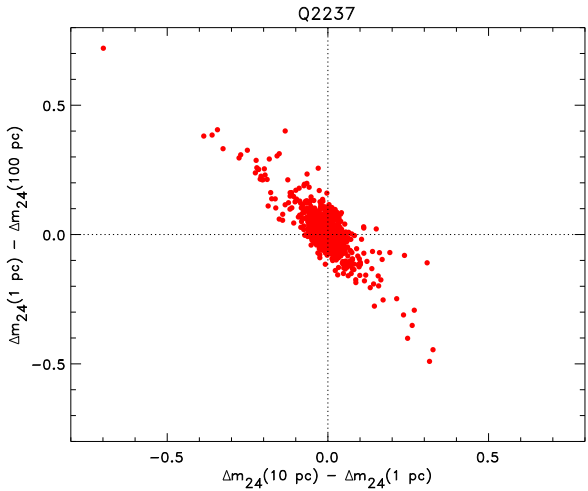} 
\vspace{1in}
\caption[]{\footnotesize Same as in Figure~\ref{fig:relmag_1}, but 
for images B and image D of Q2237+030.  In this case one quarter of
the realizations have $|\Delta m_{24}(10\pc)- \Delta m_{24}(1\pc)| >
0.05\mag$ and one quarter of them have $|\Delta m_{24}(1\pc)- \Delta
m_{24}(100\pc)| > 0.07\mag$.
\label{fig:relmag_2}}
\end{center}
\end{figure}

\begin{figure}
\begin{center}
\vspace{1.7in}
\includegraphics{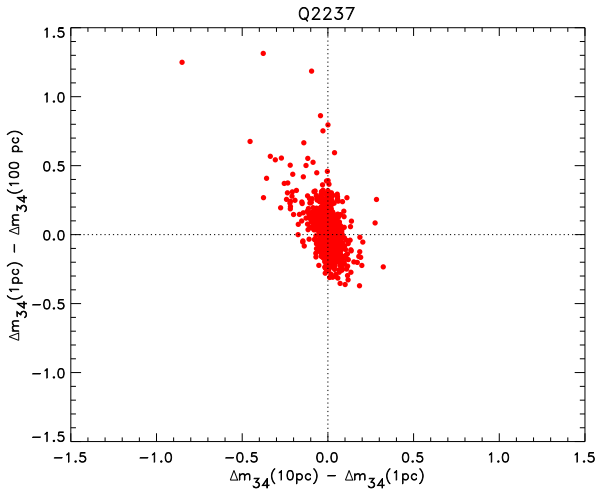} 
\vspace{1in}
\caption[]{\footnotesize Same as in Figure~\ref{fig:relmag_1} but for 
images C and image D of Q2237+030.  In this case one quarter of the
realizations have $|\Delta m_{34}(10\pc)- \Delta m_{34}(1\pc)| >
0.05\mag$ and one quarter of them have $|\Delta m_{34}(1\pc)- \Delta
m_{34}(100\pc)| > 0.17\mag$.
\label{fig:relmag_3}}
\end{center}
\end{figure}

\begin{figure}
\begin{center}
\vspace{1.7in}
\includegraphics{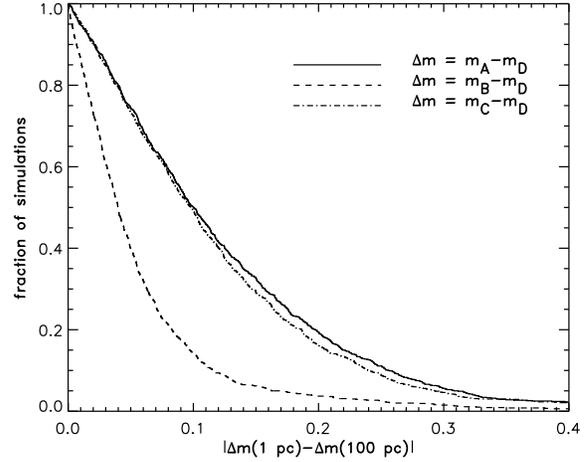} 
\vspace{1in}
\caption[]{\footnotesize Here is a histogram the cumulative
distribution of the differential magnification for the three
independent ratios in Q2237+030.  On the y-axis is the fraction of
simulations out of 1\,000.  The source sizes used are 1\,pc and
100\,pc to reflect the approximate sizes of the BLR and the NLR,
respectively.
\label{fig:stats_source_size}}
\end{center}
\end{figure}

Figures~\ref{fig:relmag_1} through \ref{fig:relmag_3} show simulations
of all three magnification ratios for Q2237+030.  In each case, three
different source sizes are used.  If the lens were smooth all three
magnification ratios would be the same and all the dots in the figures
would be located precisely at the origins in these plots.
Figure~\ref{fig:stats_source_size} shows the cumulative distribution
of the simulated differential magnification ratios.  We can see that
in half of the cases the differential magnification ratio is larger
than 0.1~mag.  If this level of measurement accuracy can be attained, 
only 1--3 lens systems will be required to rule out this level of
substructure.

Our model for Q2237+030 predicts magnifications that are relatively
modest ($\mu_A=2.28$, $\mu_B=-0.939$, $\mu_C=2.047$, $\mu_D=-2.005$).
Since the substructure has an influence on the magnification that is
proportional to the smooth components magnification ($\delta\mu\propto
\mu_o^2$), we expect that these results are fairly conservative.  Some
published models for other quad lenses predict magnifications as large
as 100.

We can see from these simulations that it is possible to detect
substructure, or intergalactic dark haloes, in a way that is
independent of any lens model, if we can measure the flux ratios to
$\sim 0.1$~mag for sources that differ in size by a factor of ten and
are centered on the same object.  This can be done with a very small
number of lens systems.  Interpreting the results in terms of the
fraction of mass in small mass haloes and the slope of the mass
function will require some modeling of the smooth lens component, but
even then the results will be much less model dependent than the more
traditional method based on single broad-band measurements of the flux
ratios.

\section{The Spectroscopic Approach}\label{sec:spec} 

\subsection{the internal structure of QSOs -- a sketch}
\label{subsec:qsos} 

In QSO spectra, the `narrow line region' (NLR) emission lines
(e.g. \oiiilam) do not vary in flux significantly over the time span
of several years \citep{kaspi:96b}.  For this reason, they have been
used for calibration in long-term spectroscopic monitoring
observations of low redshift QSOs for reverberation mapping
experiments, by which the time delays between features of different
ionization states may in principle be used to map all the components
that make up QSOs \citep{peterson:93}.  Photo-ionization arguments and
cases where the NLR-emitting region is imaged directly, indicate that
the size of the NLR extends out to 100-1000\,pc.

In the full unification picture for active galactic nuclei (AGN)
including QSOs, the variability timescales of different spectroscopic
features, and the time delays between them, are used to map out the
size and structure of different components with very different
characteristics \citep[c.f.][]{antonucci:93}.  The basic idea consists
of a supermassive black hole (M$_{bh}>10^{6}$M$_{\odot}$) in the core,
surrounded by an accretion disc, which produces the nearly
flat-spectrum continuum light seen.  The continuum flux varies on very
short timescales, less than a day, and so must be very compact (around
100\,AU or $\sim5\times10^{-4}$\,pc).  It is also known to be
microlensed in the case of Q3327+030 which puts an upper limit of
2\,000\,AU on its size \citep[][and references therein]{yonehara:01}.
Beyond that, there are permitted lines such as the Balmer-series lines
that are very broad ($v_{\rm FWHM}>10^3$\,km\,s$^{-1}$) due to
gravitationally-induced motions.  The characteristic size of this
`broad line region' (BLR) apparently scales with the intrinsic
luminosity of the host QSO (and therefore with the mass of the central
black hole; \citealt{kaspi:96a, wandel:99}), such that in luminous
QSOs (as opposed to, e.g., Seyfert~1's), the size of the BLR is on the
order of six light-months (or $\sim0.3$\,pc).  According to long-term
studies of QSOs, the typical (rest-frame) flux variations in the QSO
continuum are on the order of 10--70\% (though occasional fluctuations
as high as $\sim50\times$ are possible; \citealt{ulrich:97}), while
the BLR variations are smaller by a factor of 2--4 \citep{maoz:94}.
In the observed frame, these timescales are stretched by an additional
factor of (1+$z_s$).

\begin{figure}
\begin{center}
\vspace{3.7in}
\includegraphics{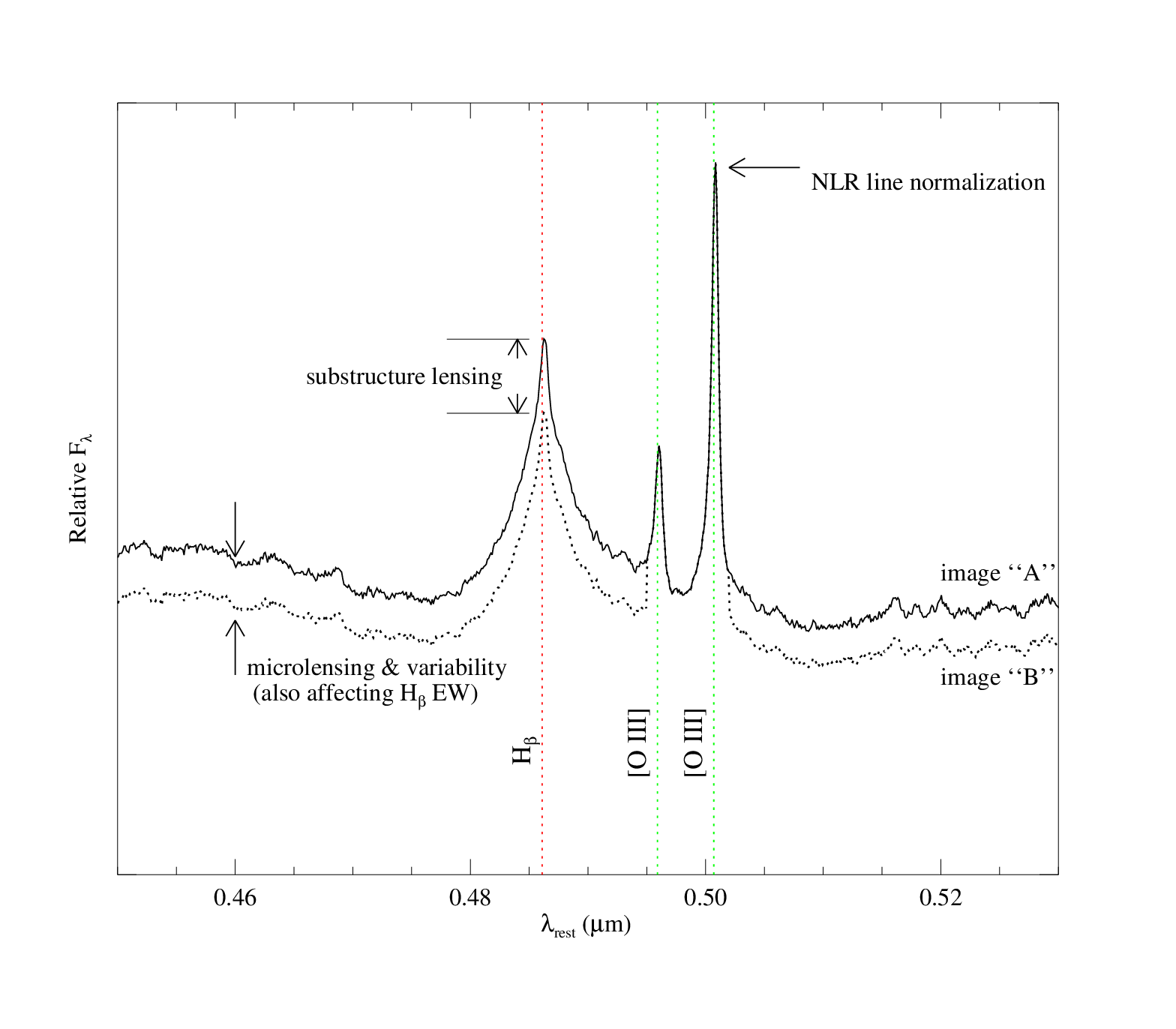}
\vspace{-1in}
\caption[]{\footnotesize In the rest-optical, the shown
rest-wavelength region (from the SDSS QSO template,
\citealt{vandenberk:01}), includes the NLR \oiiilam\ and the BLR \hb\
emission features.  Where the NLR emission line is used for
normalising the spectra of each lensed image, any residual in the
relative intensity of the BLR \hb\ emission lines after accounting for
the relative magnification in the relevant gravitational lens model,
will most likely be due to compound-lensing due to substructure in the
foreground `smooth' lens gravitational potential, whereas intensity
variations in the continuum will be due to a combination of
substructure- and micro-lensing, and other effects as discussed in the
text.}\label{fig:pqso}
\end{center}
\end{figure}

There is some evidence for anti-correlations between the
rest-ultraviolet continuum flux and the equivalent width of certain
emission lines (\citealt{baldwin:77}; \citealt{green:01}), which may
also be related to the correlation between the central black hole mass
and the continuum flux \citep{wandel:99}.  The general trend claimed,
is that the more luminous QSOs have less pronounced broad-line
emission line components, the \oiii/\hb\ central-flux ratio is small,
and the rest-ultraviolet and -optical \feii\ multiplet emission lines
are more prominent, obfuscating the `true' local continuum level that
underlies these emission lines.  There is strong (brightening)
evolution with redshift observed in the luminosity function of QSOs,
so that in general, high-redshift optically-selected QSOs are less
likely to have strong \oiii\ emission.  The QSO templates from the
Large Bright QSO Survey (LBQS; \citealt{francis:92}) and the Sloan
Digital Sky Survey (SDSS; \citealt{vandenberk:01}) are compiled from
objects over a broad range of redshifts, relatively few at $z<1$
(where the rest-optical window around \hb\ is still visible).
Therefore, the \hb\ region in these templates is drawn from objects
that are relatively local and of lower luminosity, and so generally
show a misleadingly large \oiii/\hb\ ratio, if they are to be applied
to higher-redshift considerations.

The only definite way to determine the characteristics of a particular
system is by direct observations in the rest-optical.  For example, in
the case of B1422$+$231 at $z_s=3.62$, \citet{murayama:99} find that
the \feiilam\ emission is relatively weak compared to \hb, similar to
lower-redshift QSOs.

\subsection{the spectroscopic approach for substructure}
\label{subsec:specapproach} 

Consider a case where the spectrum of all images in an Einstein Cross
are obtained, in a wavelength range where both NLR and BLR emission
features are visible at high signal-to-noise (SNR\,$>\!\!>$\,10), and
with sufficient spectral resolution to clearly resolve the shapes of
the lines ($v_{res}$\ltsima$200$\,km\,s$^{-1}$).  The NLR emission
lines are used to normalise the spectra to each other, and we wish to
attempt to disentangle the substructure signature from other effects.
There are several systematic errors to consider.

\medskip{\bf Differential Reddening --} As each lensed image follows a
different line of sight through the lensing galaxy, if the gas and
dust in the galaxy are sufficiently patchy or position dependent, it
is possible that there will be differential reddening that may affect
the measured continuum slope and relative emission line fluxes.
\citet{falco:99} measured the $\Delta\,E(B-V)$ in 23 lens galaxies
using HST broadband colours, under the (optimistic) assumption that
variability and microlensing were not important.  In recent
broadband-magnitude-based substructure estimates, the colour
corrections estimated by the above work were applied, which, in light
of the possibility of microlensing, may introduce unquantifiably
systematic errors for any one system.  In the spectroscopic approach,
choosing NLR and BLR emission lines that are close in wavelength, the
differential reddening is very small.  Taking the median value of
$\Delta\,E(B-V)=0.04$\,mag in the \citet{falco:99} optically-selected
sample, and assuming a MWG-type extinction law \citep{cardelli:89}, we
estimate that flux ratio between \oiiilam\ and \hb\ for $z_s=3$ and
$z_l=1$ will vary by less than 0.07\% and can therefore be ignored.
For comparison, between \oiilam\ and \mgiilam\ the variation will be
about 1.4\%.

\medskip{\bf Microlensing --} Some microlensing by stars probably
occurs in most lenses, but it is most clearly seen in quadruple lenses
\citep*[][and citations thereof]{witt:95}. Microlensing events last
months or years, much longer than the time delay between images, but
(especially at caustic crossings, which can occur quickly) can magnify
an image by as much as a whole magnitude \citep*[e.g.][]{witt:95}.
For super-microarcsec sources, like the BLR of a QSO, stellar
microlensing is insignificant (Figure~\ref{fig:mcls}), but for smaller
sources, like the more compact QSO non-stellar continuum, microlensing
will affect the magnification. Hence, observations of the relative
flux of the broad component of BLR lines in the images (e.g. after any
narrow component has been fit out) will reveal the effects of lensing
by CDM substructure alone, while their {\em equivalent widths} can be
used to probe ongoing microlensing because they are measured relative
to the continuum \citep{mcleod:98}.  Thus, in a spectroscopic
experiment as we have described, it would be possible to place a limit
on microlensing and substructure lensing simultaneously.

If the QSO accretion disc has a steep temperature gradient, it is also
possible for microlensing of the disc to produce {\em
wavelength-dependent} magnification in the continuum of the affected
image (as seen in HE\,$1104$--$1805$ by \citealt{courbin:00}).  This
does not affect the BLR emission line fluxes, as compared against NLR
lines, though it is also spectacular evidence for microlensing.

\medskip{\bf Intrinsic QSO Variability -- } The continuum flux
variations in QSOs can be quite large and rapid, on timescales
comparable to the time delay between images.  The BLR emission line
fluxes vary less dramatically, and only reflect the more pronounced
fluctuations seen in the continuum, with a several-month time-lag in
the rest frame.  For source redshifts $z_s\approx2$, the
observed-frame time-lag becomes close to a year.  As long as
variations in the observed frame happen on timescales that are
significantly longer than the time delay between images in a given
lens system, using the BLR lines as a substructure probe should be
robust.

\section{Discussion}\label{sec:disc}

The ideal target is a four-image lensed QSO with an exceptionally
well-determined (smooth) gravitational lens model, at a redshift such
that the ideal \hb/\oiii\ features are redshifted into convenient
observable windows, unobscured by strong atmospheric emission bands.
It is also desirable for the target QSO to have a high \oiii/\hb\
ratio, which typically also means that the \hb\ emission will be
dominated by its BLR component, and also that the \feii\ complex
shortward of \hb\ will be weak, such that the underlying local
continuum can be measured securely.  The redshifts of the sources in
most Einstein Crosses are known; their \hb/\oiii\ features have only
been measured, at low signal-to-noise, in very few cases.  Therefore,
in many cases exploratory observations may be necessary.  Several
systems will need to be studied in detail in any case, to build up the
statistics needed for a definitive measurement.

As illustrated in Figure~\ref{fig:pqso}, in the rest-optical the
\oiiilam\ and \hb\ lines are close, but separated enough to be
distinguishable at even moderate spectral resolution.  At the
redshifts of most gravitationally lensed QSOs (typically $z>2$), these
features are shifted into one of the near-infrared windows at
$\lambda>1\mu$m ($JHK$).

In the Appendix we investigate the expected statistical uncertainties
in a measurement the ratios between lines and continuum.  Based on a
`typical' QSO (e.g the SDSS template; \citealt{vandenberk:01}), the
BLR to NLR emission-line ratio may be measured with an error of
$\sigma_\omega \simeq 0.3/SNR$, where $SNR$ signal to noise ratio per
resolution element with which the continuum is measured.  Likewise,
the continuum to NLR emission-line ratio can be measured with error
$\sigma_c \simeq 10/SNR$.  If in the continuum a $SNR\sim 100$ is
achieved, then based on our simulations in
\S\ref{sec:lensing-simulations} (with $\sim 10\%$ of the surface
density in compact objects below $\sim 10^6\msun$, as seems to be the
case in $\Lambda$CDM), then we expect the line ratios between the
images of a typical lens system to differ by an amount {\em larger}
than the measurement error.  The situation is actually a bit better
than this because all the images can be used to fit the full line
shapes simultaneously.  Observations of additional lens system can be
used to improve the sensitivity to the substructure mass fraction.

These levels of the SNR are easily feasible with 8-m class telescopes
for many Einstein Crosses and relatively moderate amounts of observing
time.  For instance, the Einstein Cross Q2237+030 QSO images, at
$z_s=1.69$, have infrared magnitudes of $H\approx16.0$ (from
CASTLES\footnote{The CfA-Arizona Space Telescope LEns Survey of
gravitational lenses, at {\tt cfa-www.harvard.edu/castles/}}).  With 3
hours of integration on an 8-m telescope with
$v_{res}$\ltsima$50$\,km\,s$^{-1}$, the relevant part of the
continuum, will be detected with $SNR$\gtsima$100$.

\section{Conclusions}\label{sec:conc}

In $\Lambda$CDM, a significant fraction of the dark matter in
$L\sim\,L_*$ galaxy haloes is in coherent substructure, with masses of
$10^4$--$10^8$\,M$_{\odot}$, which does not seem to have baryonic, or
at any rate luminous, counterparts.  The substructure and small-scale
intergalactic structure should be detectable by its gravitational
signature in multiply-imaged QSOs.  If such substructure is
sufficiently abundant along the images' lines of sight, the light from
different emission regions of a QSO will be magnified differently.
This would be a model independent signature of small scale
substructure.

Broadband photometry is primarily a measurement of the continuum flux,
which is affected by rapid fluctuations intrinsic to the QSO, and by
microlensing caused by stars.  The magnitudes of both of these effects
are at least comparably to what substructure is expected to produce,
and conventionally can only be tracked by extensive monitoring.

We propose that these fundamental limitations can be largely overcome
by obtaining high signal to noise spectra ($SNR\gsim 100$) of two or
all lensed images simultaneously, that include prominent NLR and BLR
features.  The NLR features are used to normalise the spectra to each
other, and then the relative fluxes of the BLR line(s) should be
primarily due to substructure differential magnification.  Intrinsic
variability in BLR lines is observed to amount to $\sim10$--$35$\%,
but relatively smoothly, and on characteristic timescales of greater
than six months in the rest frame (and 1+$z_s$ times longer in our
observed frame).  Compared to the time delay of only days or weeks
between the images in typical Einstein Crosses, the {\em intrinsic}
BLR fluxes can be assumed to be constant.

The ideal setup is with an integral field unit (IFU) in the
near-infrared (NIR), targeting the rest-optical \oiii\ \& \hb\
emission lines.  With the IFU setup, it is possible to get
high-quality spectra of all images (and possibly of the lens, as
well), simultaneously.  The alternative of a set of longslit
observations can be made more efficient by aligning on pairs of lens
images, positioning the slit close to the parallactic angle to
minimize the effects of differential refraction.  Efficient
implementation will require an 8-m class telescope, with excellent
seeing conditions ($\sim0.6''$), perhaps with adaptive optics.

This experiment will provide a concrete measurement, or at least place
a severe upper limit on the fraction of dark-matter substructure in
galaxy haloes and intergalactic space as predicted by the popular CDM
models of structure formation.  These models are presently in a state
of crisis because of the lack of observations of this substructure.

\section*{Acknowledgments}

We thank Piero Madau, Andrew Blain and Aaron Barth for useful
discussions.  LAM was supported by the PPARC Rolling Grant
PPA/G/O/2001/00017 at the University of Oxford, and thanks the
Institute of Astronomy at the University of Cambridge for their
hospitality during the course of this work.  We gratefully acknowledge
the CASTLES project for providing such a useful resource to the
community.

\bibliographystyle{mn2e}

\clearpage
\renewcommand{\theequation}{A\arabic{equation}}
\setcounter{equation}{0}  
\renewcommand{\thefigure}{A\arabic{figure}}
\setcounter{figure}{0}  
\onecolumn
\section*{\centerline{APPENDIX}}
\section*{\centerline{Expected Uncertainties in line strength ratios}}

We wish to estimate how well the ratio of the strengths of the broad
and narrow lines can be measured.  To do this we will model the flux
in the $i$th resolution of the spectrum as consisting of four
contributions: the broad lines ($\omega' W^b_i(\Delta_b)$), the narrow
lines ($f^n W^n_i(\Delta_n)$), the continuum emission ($c' C_i$) and
instrumental noise ($N_i$).  The functions $W^b_i$, $W^n_i$ and $C_i$
are normalised so that their sum over all resolution elements is one
and their widths (and perhaps some other shape parameters) are
$\Delta_b$ and $\Delta_n$.  We are interested in the ratio
$\omega\equiv \omega'/f^n$.  The flux is then
\begin{equation}
F_i = I_i f^n (\omega W^b_i(\Delta_b) + W^n_i(\Delta_n) + c C_i) + N_i
\end{equation}
where $I_i$ is the extinction and $c=c'/f^n$.  We will rewrite the
extinction curve as $I_i f^n=a h_i$ with the normalisation $\sum_i h_i
W^n_i =1$ so that $a$ is the total flux in the narrow lines after
extinction correction.  If we assume Gaussian, uncorrelated noise the
log of the likelihood function will be
\begin{equation}\label{likelihood}
-\ln{\mathcal L}= \sum_i \frac{1}{2\sigma_i^2} \left[ a h_i (\omega
 W^b_i(\Delta_b) + W^n_i(\Delta_n) + c C_i) - F_i \right]^2 +
 \ln(2\pi\sigma_i^2)/2.
\end{equation}
For our present purposes it is convenient to replace the sum over
resolution elements in (\ref{likelihood}) with an integral over
wavelength by making the substitution $\sum_i\rightarrow
\delta\lambda^{-1}\int d\lambda$ where  $\delta\lambda$ is the width in
wavelength of the resolution elements.  We will assume that the noise is
constant over the wavelength range considered and given by $\sigma_n$.

The Fisher information matrix is defined as
\begin{equation}
F_{p_1p_2}=- \left\langle \frac{\partial^2\ln{\mathcal L}}{\partial
p_1\partial p_2} \right\rangle
\end{equation}
where $p_1$ and $p_2$ are two parameters that are to be constrained by
the measurements.  The variance of the minimum variance unbiased
estimator of a parameter $p$ is given by the Cramer-Rao bound
\begin{equation}
\sigma^2_p = (F^{-1})_{pp}.
\end{equation}
To calculate this we must identify the parameters in
(\ref{likelihood}) that are to be fit to the spectrum.  The region of
the spectrum we are interested in, shown in Figure~\ref{fig:pqso}, is
relatively small compared to the complete spectrum and it is in region
where the continuum has a shallow minimum in template QSO
spectra\citep{vandenberk:01}.  For simplicity we will approximate the
extinction, $h(\lambda)$, and continuum, $C(\lambda)$, as constant
over the region of interest, but we will allow their normalisations to
vary.  Formally, this is a valid approximation if $d\ln
C(\lambda)/d\lambda, d\ln h(\lambda)/d\lambda \ll 1/\Delta\lambda$
where $\Delta\lambda$ is the range of wavelength considered.  We will
also assume that the redshift of the QSO is well established so that
the positions of all lines are known.  This leaves five parameters to
be determined -- $\omega$, $c$, $a$, $\Delta_n$ and $\Delta_b$.  There
are then 15 independent elements in the Fisher matrix which we
calculate.  In general the profiles of the lines are functions of
multiple parameters which could be added to the list, but for
simplicity we will leave these out here and take the lines to be
Gaussian.  Once these matrix elements are calculated the matrix can be
inverted numerically.

It is convenient to express error estimates in terms of the signal to
noise per resolution element with which the continuum can be measured.
In our parameterization this is
\begin{equation}
\sigma_o\equiv \frac{\sigma_n\Delta\lambda}{ac\delta\lambda}.
\end{equation}
A noise level of $\sigma_o\sim 0.01$ may be
reasonable. Table~\ref{table} shows the estimated errors for several
choices of fiducial parameters.  The ratio of the continuum flux
within the range $\Delta\lambda$ to the narrow line flux, $c$, is
estimated from the height of the narrow lines relative to the
continuum in composite spectra \citep{vandenberk:01}.  We use only
\hb~and \oiiilam~for the estimates and assume that no lines overlap
except the BL and NL \hb.  The source is at $z_s=2$.
\begin{table}
\begin{center}
\begin{tabular}{cccccccccccc}
\multicolumn{7}{c}{Parameter Values} & \multicolumn{5}{c}{Estimated
Errors in units of $\sigma_o$} \\
$R^\spadesuit$ & $\Delta\lambda^\dagger$ & $\omega$ & $c$ & $acC(\lambda)^\ddagger$
& $\Delta_n$ & $\Delta_b$ & $\sigma_\omega$ & $\sigma_c $ & $\sigma_a/a$ &
$\sigma_{\Delta_n}/\Delta_n$ & $\sigma_{\Delta_b}/\Delta_b$ \\
\multicolumn{7}{c}{----------------------------------------------------------------------}
&\multicolumn{5}{c}{-----------------------------------------------------}\\
3000 & 50,000 & 1.0 & 22.2 & $1.7\times 10^{-16}$ & 300 & 5,000 & 0.255 & 13.1 & 0.538 & 0.531 & 1.19 \\
3000 & 100,000 & 1.0 & 19.9 & $1.7\times 10^{-16}$ & 500 & 10,000 & 0.148 & 6.88 & 0.309 & 0.308 & 0.746 \\
\end{tabular}
\end{center}
\footnotesize $^\spadesuit$ Resolution of spectrograph,
$R=\lambda_o/\Delta\lambda$. \\ $^\dagger$ $\Delta\lambda$, $\Delta_n$
and $\Delta_b$ are expressed in km/s.\\ $^\ddagger$The continuum flux
in $\mbox{erg}\mbox{ cm}^{-1}\mbox{s}^{-1}\mbox{\AA}^{-1}$.
\caption[]{\footnotesize }
\label{table}
\end{table}

In this analysis we have ignored the noise associated with subtracting
the sky and the absolute flux calibration.  This might interfere with
the measurement of the total flux from the NL, $a$, but should not
seriously interfere with measurements of the ratios between NL and BL
or between the continuum and the NL.  We also ignored the fitting of
the ratios between NLs which we do not think would affect our
estimates by much.  The line shape estimates are relatively
statistically independent from the line ratio estimates so we also do
not expect that having more complicated line shapes would greatly
change our estimates.

\end{document}